# Face-centered-cubic titanium in Ti/Al multilayer thin films synthesized by magnetron sputtering technique

*Ramaseshan Rajagopalan[#,\*], Arup Dasgupta[†], Ramachandran Divakar[†], Sitaram Dash[#], Nithya Ravindran, Saroja Saibaba[†], Ashok Kumar Tyagi, Supriya Bera[$] and Indranil Manna[$]*

[#] Surface and Nanoscience Division, [†]Physical Metallurgy Division,
Indira Gandhi Centre for Atomic Research, Kalpakkam – 603 102, India
[$]Metallurgical & Materials Engg. Dept., I.I.T., Kharagpur, W.B. 721302, India

\*Corresponding author's email id: seshan@igcar.gov.in

**ABSTRACT**

Ti/Al multilayer thin films with precise thickness have been deposited using a combination of dc and rf magnetron sputtering techniques. Cross-sectional transmission electron microscopy (TEM) revealed unmixed fifteen parallel and alternate layers of Ti and Al with sharp interfaces, each measuring 27 nm and 15 nm in thickness, respectively. The Ti layer was composed of hcp and fcc phases while the Al layer was fcc. Both x-ray diffraction (XRD) and selected area electron diffraction (SAED) analysis confirmed the identity of these phases. Detection of fcc-Ti in as-deposited the Ti/Al multilayer thin film by XRD established that the fcc-Ti phase is not an artifact of TEM sample preparation, as have been envisaged by some of the previous researchers. The fcc-Ti phase appeared when dual rf guns were used for Ti deposition and the diffraction peak intensity corresponding to fcc phase increased when the gun power was raised. A modified equation of state based thermodynamic analysis confirmed that the formation of hcp phase as opposed to the thermodynamically stable fcc phase of pure Ti is due to crystallite size reduction and not impurity driven.

# INTRODUCTION:

Artificially modulated thin film multilayer architectures are used in modern technologically important areas like giant magneto resistance (GMR), X-ray reflectivity, nuclear fuel shields and sensors. It is important to study the structural characteristics and stability of these multi-layer films for the intended use with precision and reliability. The present study pertains to Ti/Al multilayer thin films grown on glass substrates by magnetron co-sputtering technique. Ti/Al bilayers find diverse applications such as ohmic contact for InP semiconductors[1] as well as enhanced corrosion protection for steel structures[2]. Decreased thermal diffusivity in these multilayer thin films compared to their elemental films or bulk materials[3] make them suitable as barrier layers while joining titanium aluminides efficiently[4].

Elemental Ti is known to exist in two crystal structures: body centered cubic (bcc, β-Ti, A2) at high temperature (>882ºC) and hexagonal close packed (hcp, α-Ti, A3) at low temperature. Shechtman *et al*[5] and van Heerden *et al*[6] first reported the occurrence of fcc-Ti in sputter deposited Ti-Al multi layer thin films by TEM. Subsequently, Banerjee *et al.*[7] reported the presence of a metastable fcc-Ti phase in similar Ti-Al multilayer films and attributed the phenomenon to size dependent phase formation. It is known that physical vapor deposition (PVD) techniques such as magnetron sputtering is a non-equilibrium process[5-6] in which metastable phases can nucleate and grow at room temperature. On the contrary to the above hypotheses, Bonevich *et al.*[8] argued that the formation of fcc-Ti was purely an artifact of TEM specimen preparation method rather than an intrinsic property of the synthesized material. This inference was based on identification of the metastable fcc-Ti phase only from analysis of the SAED patterns while no evidence for this phase was obtained from XRD of the as deposited material. Banerjee *et al*[9] have reported that this metastable fcc-Ti in Ti-Al multilayers is stable only when the thickness of the individual layer is less than 10.5 nm.

Another group of researchers[10] has reported that Al influences the formation of fcc-Ti up to about 5.5 monolayers which amounts to only about 1.2 nm assuming a maximum inter-planar spacing of 0.22nm. Some researchers[11] also believe that the strain energy required to overcome the energy difference between fcc and hcp crystal structures is supplied by fcc-Al layer whose lattice parameter closely matches with that of fcc-Ti (fcc-Al: $a$=0.4050 nm and fcc-Ti: $a$=0.4420 nm)[5]. It is interesting to note that the metastable fcc-Ti phase is observed not just in thin films, but Manna *et al*[12] have reported a hcp to metastable fcc polymorphic transformation in elemental titanium induced by high-energy mechanical attrition in a planetary ball mill. These authors reported that, structural instability due to negative hydrostatic pressure arising out of the nanocrystallization during mechanical attrition, lattice expansion, and high plastic strain/strain rate, are responsible for this polymorphic transformation in titanium.

Thus it appears that the metastable fcc-Ti phase in Ti/Al multilayer thin films has essentially been identified using electron diffraction technique, which some groups believe is responsible for the formation of an impurity driven artifact (TiH)[13] produced during thin foil preparation. On the other hand, this paper unequivocally establishes the presence of fcc-Ti in the as deposited Ti/Al multilayer thin film, using glancing incidence X-ray diffraction (GIXRD) technique, which is free from the possibility of introducing impurity stabilized artifact during sample preparation, as well as SAED analysis. In addition, a thermodynamic analysis is also presented to show that the concerned fcc phase formation is an internal pressure driven (negative hydrostatic pressure) change necessitated by structural instability due to nanocrystallization or size effect.

**EXPERIMENTAL:**

Ti / Al multilayer thin films were deposited on glass substrates in a high vacuum sputtering system (MECA-2000, France) using high pure Ti (99.99%) and Al (99.99%) targets and ultra pure Ar (99.999%) feed gas introduced at a flow rate of 20 sccm, at room temperature. Base pressure was typically, $6 \times 10^{-6}$ mbar and the deposition pressure was $1 \times 10^{-2}$ mbar. 300 W RF (13.56 MHz) and 400 W DC powered guns were used to deposit Ti and Al layers, respectively. Both single and dual RF guns were used for depositing Ti. The sequence of deposition has been Glass/ Al/ Ti /Al/……./Ti (30 layers). The substrate was kept at a distance of 100 mm from the target and continuously rotated at 10 rpm throughout the deposition process. Rate of sputter deposition of Al and Ti using a single gun were about 0.18 and 0.1 nm/s, respectively. While using dual guns for Ti deposition, the rate was about 0.2 nm/s. Structural characterization of the multilayer films was done using GIXRD (STOE, Germany) and various TEM models with different resolution and analytical capability (Philips CM 200, Jeol 2000EX and Tecnai F30). Cross-sectional TEM specimens were prepared by mounting the glued glass/thin film sandwiches in cross sectional configuration in hardenable resin followed by mechanical grinding, polishing dimpling and low angle Ar ion milling (BALTEC, RES101).

**RESULTS:**

Figure 1 shows the GIXRD patterns of the Ti/Al thin films deposited on glass substrates using single as well as dual guns for sputtering elemental Ti. It may be reiterated here that only a single DC gun has been used to sputter Al. The effect of increasing rf power from 200 to 300 W for the dual gun deposited Ti/Al multilayer thin film on the XRD pattern is also shown in this figure. Various XRD peaks have been identified by comparison with

JCPDS data for hcp-Ti[14] and fcc-Al[15] and marked in the figure as (hkl)$_{Ti-H}$ and (hkl)$_{Al-F}$, respectively. Interestingly, in all the XRD patterns, the strongest peak is observed from (10.0) plane of hcp-Ti as against (10.1) for polycrystalline materials taking structure factor into consideration[12]. This is indicative of preferential orientation in these films. Although conventional XRD does not provide complete information about orientation of the crystallites, it can be stated that majority of the hcp-Ti crystals are orientated such that the prismatic (10.0)$_{Ti-H}$ type of planes are parallel to the film or substrate surface (Type I) while second major set of hcp-Ti crystallites have their basal (00.2)$_{Ti-H}$ type of planes lying parallel to the substrate surface (Type II). On the other hand, Al peaks are relatively weak; one of which (111)$_{F-Al}$ overlaps with hcp-Ti (00.2)$_{Ti-H}$ while another corresponds to (220)$_{F-Al}$. Both these peaks are comparable in intensity. Since in GIXRD technique instrumental broadening ($\Delta 2\theta$) amounts to ~ 0.4 °, it is not possible to deconvolute contributions from (00.2)$_{Ti-H}$ and (111)$_{F-Al}$ from the peak at 2θ of 38.52 °. Therefore, it is not possible to comment on the degree of preferential orientation in the case of Al. As against the XRD pattern from the single gun case, the films deposited using dual guns for Ti, show an additional peak at 59.25 ° which does not match with any known reflection from hcp-Ti, bcc-Ti, fcc-Al or even their intermetallics. This peak can be attributed to a metastable fcc-Ti phase based on literature reports[17]. As a matter of fact, there are several peaks in the XRD pattern which match closely to fcc-Ti, viz., and the peak at 70.85 ° can be assigned to both hcp-Ti as well as fcc-Ti. It is also observed that as the dual rf power is increased from 200 W to 300 W, the relative intensity of the XRD peak at 59.25 ° corresponding to the fcc-Ti phase increases while another peak at 70.85 ° appears which may be attributed to the (311) of fcc-Ti[17]. This suggests that the appearance of the fcc-Ti phase is dependent on the type of gun and power used for Ti deposition. Dual gun as well as higher power increases the Ti sputtering rate resulting in an increased deposition rate of Ti which is believed to be responsible for the presence of fcc-Ti phase. It may be pointed out that Manna et al.[12] observed a similar effect of strain and strain rate in which the hcp to fcc transition was complete only in case of high strain rate deformation and not under low strain rate deformation. The crystallite size (the size of coherent crystalline domains) was calculated using the Scherrer's formula for the Ti and Al phases and found to be of the order of ~ 10 nm indicating the nanocrystalline nature of these multilayer thin films.

In order to investigate the morphology of various phases, in this multilayer system, detailed transmission electron microscopy analysis was carried out on specifically the Ti/Al multilayer thin film in which the Ti layers were deposited using dual RF guns.

Figure 2 shows the TEM bright filed image along the cross-section of the entire Ti/Al multilayer thin film deposited on glass. The glass substrate and the epoxy resin used to mount the sample in this configuration are marked in the figure. The multilayers of Ti and Al are seen as alternate dark and bright bands, respectively. The individual layers of Ti and Al were identified by Energy Dispersive X-ray Spectroscopy (EDS – not shown) using a 10nm probe. It is seen that the layers of Ti and Al are distinct and nearly parallel to the substrate surface. There are 15 layers each of Ti and Al with an average thickness of about 27 nm and 15 nm, respectively and the total film thickness is found to be 630 nm.

Figure 3 (a) shows the high resolution TEM image from a section of the Ti layer. An inset in this figure shows the Fast Fourier Transformation (FFT) of a section of the Ti lattice image. Lattice planes corresponding to $(10.0)_H$ and $(00.2)_H$ of Ti are identified from the FFT. From the combined information of the lattice image and its FFT, it is inferred that in the Ti crystallite imaged here, $(00.2)_H$ planes are aligned parallel to the substrate or the Ti/Al interface, while the $(10.0)_H$ planes are normal to it. Based on this inference, schematic of an hcp-Ti unit cell is shown overlay on the Ti lattice image. Therefore, it can be said here that the Ti crystallite shown in this figure is of the Type II (result of Fig. 1). Secondly, the crystallite is one of those rare ones which extend for about 40nm in size which is much larger than its average size of ~10 nm. However, fcc-Ti was not detected in this location. Figure 3(b) on the other hand shows the high resolution TEM image from a section of the Al layer. Barely any lattice plane is seen here because most of the Al layer has oxidised during handling of the TEM sample resulting in an amorphous aluminum oxide layer on top. This is because the free energy of formation of aluminum oxide at room temperature is only about -1020 kJ/mol of oxygen as compared to 860 kJ/mol for Ti (Ellingham Diagram).

For precise identification of the fcc-Ti phase, SAD patterns were recorded and analyzed from various locations on the cross-section of the Ti/Al multilayer. Figure 4 shows the SAD pattern from one of such regions (inset). The corresponding intensity pattern as a function of lattice spacing 'd' is shown in the figure. As in the case of figure 1, here too the SAD spectrum was compared with the JCPDS data [14, 15] for hcp-Ti and fcc-Al and literature reports[16] for fcc-Ti. Intensity peaks corresponding to different planes of hcp-Ti, fcc-Ti and fcc-Al are clearly identified. Therefore an fcc-Ti phase is found to be occurring in the region from which the SAD pattern was obtained which was further analysed and discussed below.

Figure 5(a) shows a bright filed TEM image of the region on the Ti/Al multilayer from which the SAD pattern shown in Fig. 4 was taken. As usual, alternate bright (Al) and dark (Ti) layers are seen. Dark field imaging was independently carried out on this region using first the diffraction spot corresponding to (220) of fcc-Ti (marked as ① in Fig.4) and then (00.2) of hcp-Ti (marked as ② in Fig.4). These dark field images were coloured green and blue, respectively and merged into one figure which is shown as Figure 5(b). In this manner, the regions corresponding to fcc and hcp phases of Ti can be located. However, there is one problem. The diffraction spot corresponding to (220) of fcc-Ti is very close (~ 0.01nm) to that from (220) of fcc-Al. Similarly, diffraction spot corresponding to (00.2) of hcp-Ti overlaps with that from (111) of fcc-Al. As a result certain areas in Al layers also are coloured blue or green. However, by comparing with the BF image in Fig. 5(a) the layers of Ti and Al can easily be demarcated since no intermixing of the layers has been observed in earlier figures. Hence by concentrating on the Ti layers alone, the hcp and fcc Ti phases are clearly identified. Therefore, under the dual beam deposition condition for Ti, the Ti layers are composed of a hcp-Ti and fcc-Ti, both of which are stable at room temperature. The Al layers however bear a fcc crystal structure.

**DISCUSSION:**

From the above the following are the salient observations:

1. There is tendency of preferential orientation of the Ti nanocrystalline grains in these multilayer thin films under all conditions of depositions that are discussed in this article, such as the number of guns used for Ti depositions or the power used.

2. An fcc-Ti phase, which is stable in the film at room temperature and is certainly not an artifact of TEM sample preparation as its signatures were observed in XRD as well, forms alongside the hcp-Ti phase only when more than one rf guns are used for its deposition. The phase gains in relative strength as the power of dual guns are increased.

The tendency for preferential orientation is an attribute of the deposition process which is in agreement with the work of Mahieu *et al* [17]. However, formation of a thermodynamically non-equilibrium phase of fcc-Ti at room temperature needs to be rationalized. The PVD process used in this case (Magnetron Sputtering) is anyway a non-equilibrium process in which a vapour is directly condensed into its solid phase. The departure from equilibrium for material synthesis by such vapour condensation process was reported in the literature to be as high as 160 kJ/mol, which is second to none. Therefore, it is not really surprising that even otherwise metastable phases are stabilised. Yet, this fcc-Ti phase does not form when a single rf gun is used for its deposition. Based on this important observation, the following mechanism for its formation is proposed. The PVD process can be thought to be a process in which the sputtered atoms (ad-atoms) arrive on the substrate surface at a certain rate governed by the sputtering yield at the given power applied to the gun. These adatoms are expected to laterally diffuse on the substrate surface (surface diffusion), the coefficient of which is governed by the substrate temperature until they encounter nucleation site, such as an artifact on the surface, or another atom or atoms which have already nucleated. At this site, the adatom requires a certain relaxation time during which it can readjust itself so as to match the stacking sequence. Subsequently these nuclei grow in size. In this case, the deposition was carried out at room temperature; hence the surface diffusion coefficient is anyway low resulting in random nucleation which explains the nano-crystalline sizes of the phases. When two guns are used for deposition of Ti, the arrival rate of the depositing species (Ti atoms) is nearly double as compared to that when only a single gun is used. Therefore, assuming a constant surface diffusion coefficient of the precursors on the depositing surface, the diffusion distances are drastically reduced. In other words, not all of the Ti atoms get the relaxation time that is necessary to form an equilibrium phase (i.e. hcp-Ti). A stacking error can result in a fcc phase, which has the same packing fraction as hcp. But fig. 5 showed that within a Ti layer, there is no mixture of the hcp and fcc phase along the growth direction, it is either hcp or fcc. In other words, fcc phase has not grown on hcp or vice versa. This led to the belief that the fcc-Ti phase is probably stabilized, by the underlying fcc-Al with which it has a lattice mismatch of only about 9%. This may be understood as follows: the usual ABAB stacking of the hcp crystal structure can be faulted to ABCABC…. kind of crystal structure as in the case fcc lattice under the influence of Al. But the stacking fault energy of hcp Ti is considered to be very high (>300 mJ/m$^2$) [16]. Recently Ghosh et al. [19] showed that in Ti base

Al systems, the stacking fault energy of hcp-Ti is greatly reduced by Al addition. Hence it is not unlikely that during the initial stages of Ti growth when Al coverage is much higher than Ti, an fcc stacking in Ti as in Al is promoted. In short, we can say that under the joint influence of high arrival rate Ti adatoms and the Al underlayer, the fcc-Ti phase is stabilised. The resultant product is either hcp or fcc-Ti on Al.

In order to further investigate the thermodynamic genesis of the said fcc phase formation, an analytical exercise was carried out based on the lattice instability model. The model uses an isothermal equation of state developed by Rose et al.[18], which was subsequently adopted by Fecht[20] to calculate the hydrostatic pressure ($P_h$) at the grain boundaries in nanocrystalline materials. $P_h$ can be calculated using the equation:

$$P_h = -\frac{3B_0[(V/V_0)^{\frac{1}{3}} - 1]}{(V/V_0)^{\frac{2}{3}}} \exp(-a^*)(1 - 0.15a^* + 0.05a^{*2}) \quad \ldots\ldots\ldots\ldots\ldots (1)$$

where $B_o$ is the equilibrium bulk modulus, $V_o$ is the volume per atom of the equilibrium coarse-grained material, $V$ is the volume per atom in the nanocrystalline state, and $a^*$ is the scaling parameter defined as $a^* = (r_{WS} - r_{WSe})/l$, where $r_{WS}$ is the real Wigner–Seitz cell radius of the atom with a volume $V$, $r_{WSe}$ is the Wigner–Seitz cell radius of the atom with a volume $V_o$, and $l$ is the length scale characteristic of the metal. The terms $V$ and $V_o$ can be calculated from the free volume per atom ($\Delta V_F$), where $\Delta V_F = (V-V_o)/V_o$.

The crystallite size for the Ti phase was obtained using Scherrer formula (Fig. 1) and found to be ~10 nm. The lattice parameter ($a$) of the fcc Ti phase is determined from the XRD pattern as $a = 0.4406$ nm. Accordingly, $V_o$ in the fcc unit cell is calculated to be 0.0214 nm$^3$/atom. In comparison, $V_o$ for hcp-titanium is 0.0177 nm$^3$/atom,[20] Thus, the fcc phase formation seems to be associated with large ($\approx 21\%$) increase in volume per atom ($V_o$). Significant reduction in crystallite size, $d_c$ and increase in $V_o$ (or molar volume, $V_m$) may lead to a substantial increase in Gibbs energy ($\Delta G$) in ultrafine crystals through the Gibbs-

Thompson equation: $\Delta G = 4\gamma V_m/d_c$, where $\gamma$ is the interfacial energy[21]. This increase in $\Delta G$ may eventually cause a high order of lattice instability to compel the incoming Ti atoms to adopt a polymorph other than the hcp-Ti with greater topological or surface density of atoms without altering the crystal packing density (= 74 %). In this regard, it is established that continued reduction in $d_c$ near 10 nm leads to a substantial increase in negative (from core to periphery) hydrostatic pressure ($-P_h$) on the grain boundary of nanocrystalline Ti which eventually stabilizes the fcc[12] phase. Adopting the same model,[18, 20, 12] the magnitude of free volume per atom ($\Delta V_F$) is calculated to be 0.098 for $d_c$ = 10 nm. Using the above values, $-P_h$ = 3.86 GPa (Eqn. 1). The combined effect of significant increase in $-P_h$ and $\Delta V_a$ [= $V_{a,fcc}$ − $V_{a,hcp}$ = 0.0214–0.0177 = 0.0037 nm$^3$/atom] is equivalent to an elastic strain energy increment of $-P_h \times \Delta V_a$ = 1.43 ×10$^{-20}$ J/atom, while the theoretically predicted change in enthalpy ($\Delta H$) in the event of a fcc crystal structure formation instead of the hcp is 5.5 × 10$^{-21}$ J/atom[22]. It is found that $\Delta H$ as a result of fcc phase formation is an order of magnitude lower than the strain energy value. The increase in strain energy may account for the structural instability owing to the Gibbs-Thompson effect. This structural instability is believed to be responsible for stabilising the fcc-Ti phase.

**CONCLUSION:**

Ti/ Al multilayer thin films have been successfully deposited using magnetron sputtering technique with no evidence for intermetallic phases. TEM studies showed that the Ti and Al layers were unmixed. There were 15 layers each of Ti and Al. The average thickness of the Ti and Al layers were about 27 nm and 15 nm, respectively. Apart from hcp-Ti and fcc-Al, an fcc-Ti phase has also been identified in these films by XRD and confirmed by TEM studies. Identification by XRD showed that the fcc-Ti phase is not an artifact of TEM sample preparation but an intrinsic property of the material. The fcc-Ti phase appeared when dual guns were used for Ti deposition and whose relative strength increased when the gun power was raised. The formation of this otherwise metastable phase was understood in terms of an atomistic model in which it was believed that higher arrival rate Ti atoms on deposition surface and the influence of the underlying fcc-Al layer was responsible for its formation. The Ti phase exhibited preferential orientation such that majority of the hcp-Ti unit cells were so arranged such that their c-axis was parallel to the substrate surface/ interface.

**ACKNOWLEDGMENT:**

The authors are pleased to acknowledge Dr. Tom Mathews and Mr. R. Krishnan for their help in experimentation, Mrs. S. Kalavathi for GIXRD experiments and Mr. V. Sankara Sastry for very useful discussion.. The authors would also like to acknowledge Dr. C.S. Sundar, Dr. M. Vijayalakshmi, Dr. T. Jayakumar and Dr. Baldev Raj, for their constant encouragement and support for this work.

FIGURE CAP*TIONS:*

*Figure 1. X-ray profile of T*i/Al multilayer thin films with single and dual guns for Ti deposition along with Ti-H pattern.

Figure 2. Cross sectional TEM image of Ti/Al multilayer thin film deposited by dual gun magnetron sputtering.

Figure 3. High resolution TEM images of (a) Ti and (b) Al layers. Inset of Fig. 3(a) shows the FFT of the Ti lattice. The orientation of a hcp-Ti unit cell is schematically shown overlaid on the HR image in this figure.

Figure 4. Radial intensity profile of selected area diffraction pattern from the Ti/Al multilayer thin film showing the presence of fcc-Ti along with fcc-Al and hcp-Ti. The SAD pattern is shown in the inset.

Figure 5. (a) Bright filed image of Ti/al multilayers; (b) Dark field image showing green color for d-value of 0.15 nm (220-fcc-Ti and 220-fcc-Al) and blue color for d-value of 0.23 nm (00.2-hcp-Ti and 111-fcc-Al) and corresponding to the same region as shown in (a).

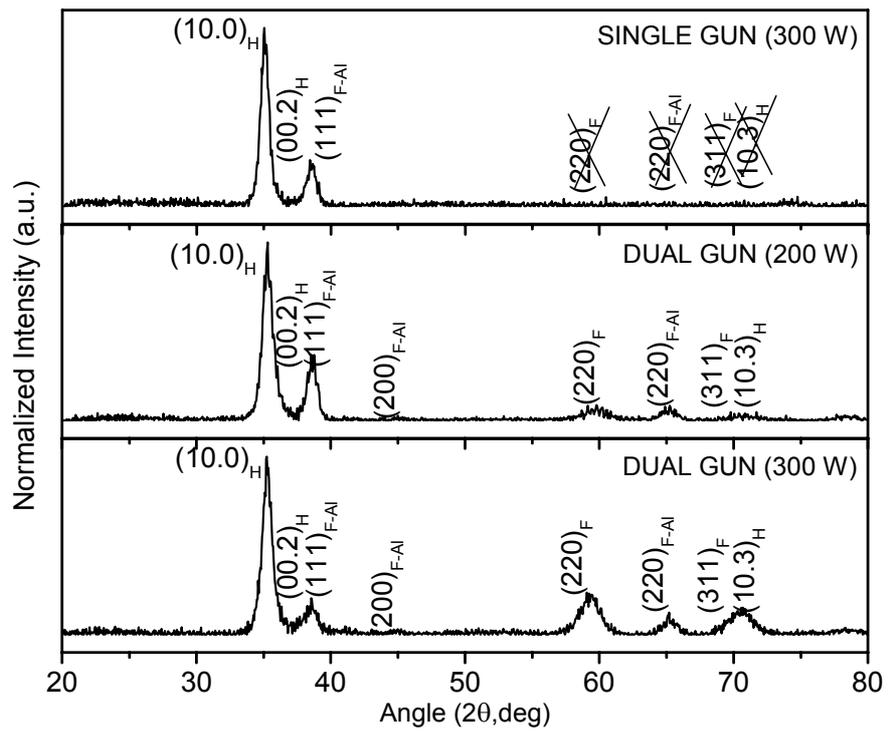

Figure 1.

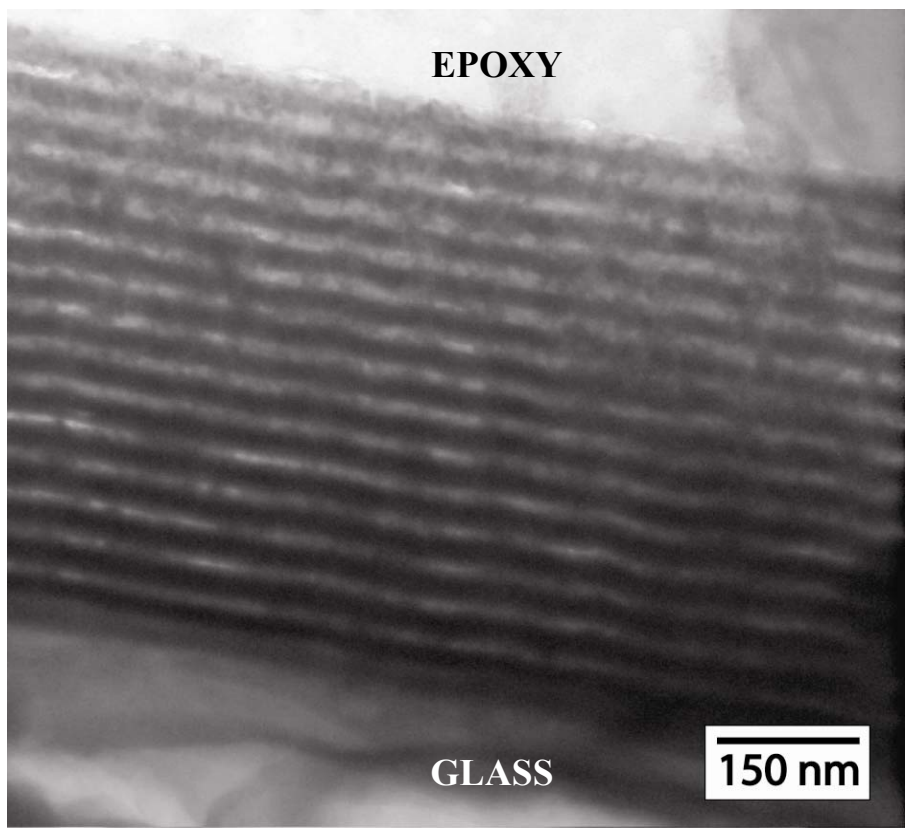

**Figure 2**.

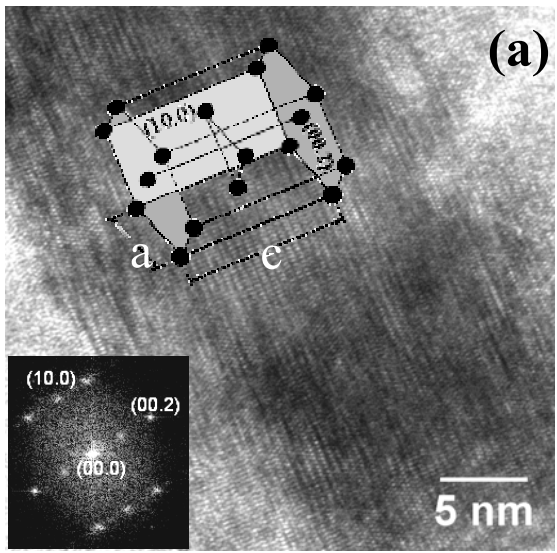 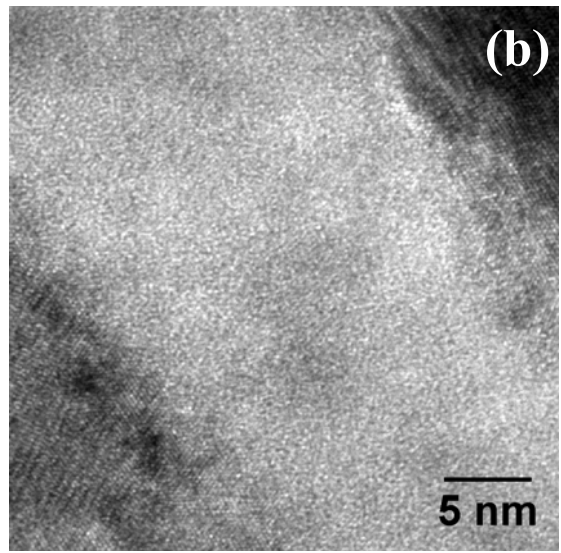

**Figure 3**.

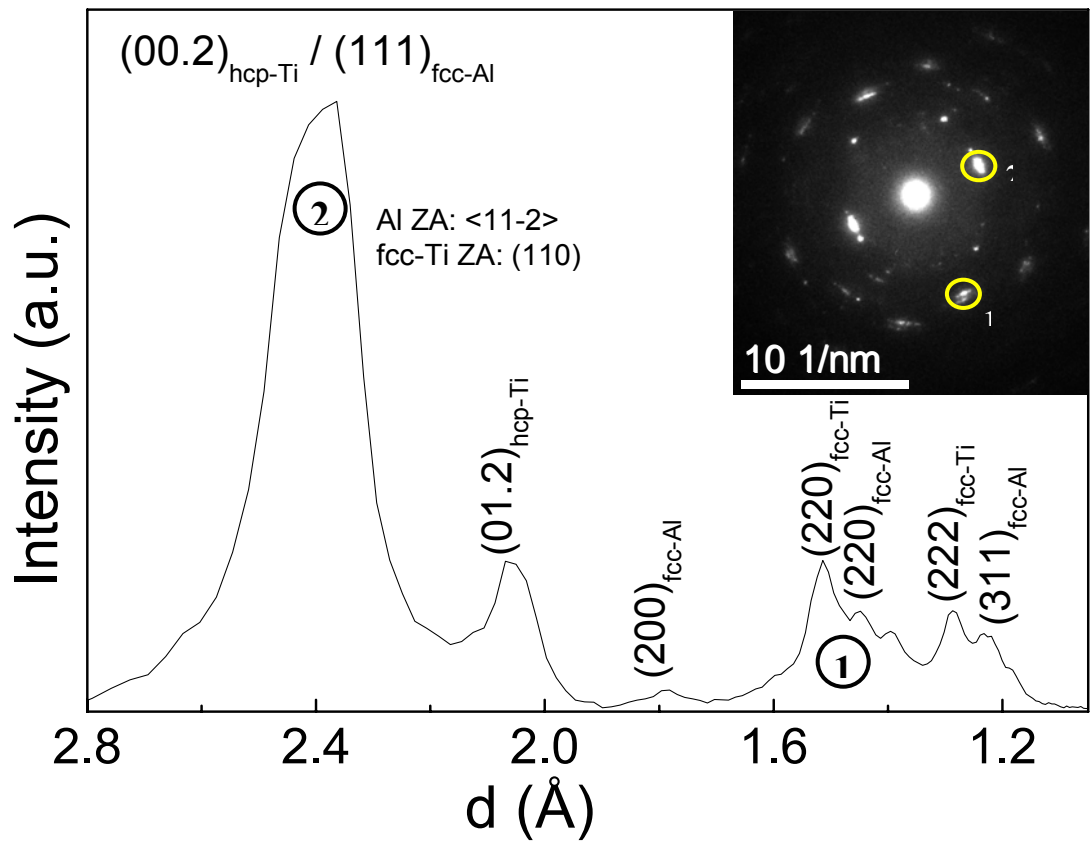

**Figure 4**.

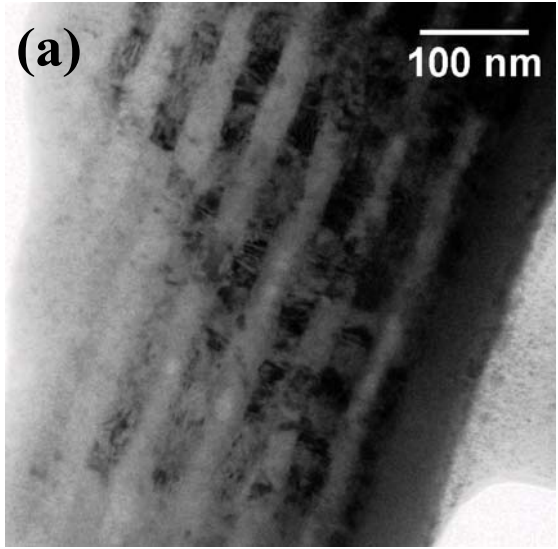 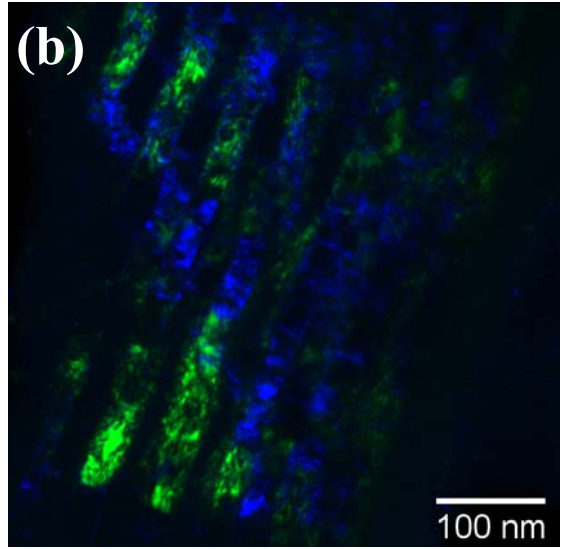

**Figure 5**.